\documentclass{80SA}
\usepackage{times}

\usepackage{amsmath, amssymb}
\usepackage{graphicx}
\usepackage{bm}

\title{
Continuous-Time Quantum Monte Carlo Study\\ of Local Non-Fermi Liquid State in the Multichannel Anderson Model
}

\author{Junya Otsuki\thanks{E-mail address: otsuki@cmpt.phys.tohoku.ac.jp} 
}
\inst{Department of Physics, Tohoku University, Sendai 980-8578 
}
\abst{%
The impurity Green's function $G_f$ in the local non-Fermi liquid state is evaluated
by means of the continuous-time quantum Monte Carlo method extended to the multichannel Anderson model.
For $N=M$ (where $N$ and $M$ are numbers of spin components and channels, respectively), 
$G_f$ is expressed as $-{\rm Im}G_f(\omega+{\rm i}0) = c - b |\omega|^{1/2}$, and the zero-frequency value $c$ depends only on $N$ ($=M$).
A corresponding impurity self-energy
at low frequencies is composed of two parts: a resonance term related to $c$, and a non-Fermi liquid term proportional to $|\omega|^{1/2}$.
The characteristic energy scale is discussed in terms of the non-Fermi liquid term in the self-energy.
}

\kword{continuous-time quantum Monte Carlo (CT-QMC), two-channel Kondo effect}

\begin{document}
\maketitle

\section{Introduction}
The multichannel Kondo effect is a typical example that leads to a local non-Fermi liquid ground state\cite{Nozieres-Blandin}.
It has been recognized that the peculiar low-temperature behaviors observed in 
uranium compounds and metals with uranium impurities
are due to the two-channel Kondo effect\cite{Cox-Zawadowski}.
This kind of non-Fermi liquid state has been investigated from a more general point of view based on models generalized to ${\rm SU}(N) \otimes {\rm SU}(M)$ symmetry\cite{Cox-Ruckenstein93}.
Then, their critical nature has been discussed extensively\cite{Ludwig-Affleck, Jerez98}.

Regarding the (single-channel) Kondo problem, the Anderson Hamiltonian gives clear insight\cite{Yamada, Yosida}:
the ground state is connected to that in the non-interacting limit.
In this analogy, the multichannel Kondo effect can be addressed based on an Anderson Hamiltonian\cite{Schiller98}.
The inclusion of the impurity charge degree of freedom enables us to describe the local dynamics via the impurity Green's function.
We thus consider the ${\rm SU}(N) \otimes {\rm SU}(M)$ multichannel Anderson model given by\cite{Cox-Zawadowski}
\begin{align}
	{\cal H} = 
	\sum_{\bm{k} \alpha \mu} 
	\epsilon_{\bm{k}} c_{\bm{k} \alpha \mu}^{\dag} c_{\bm{k} \alpha \mu}
	+ E_{\rm ex} \sum_{\alpha} X_{\alpha, \alpha} \nonumber \\
	+ V \sum_{\alpha \mu} 
	\left( X_{\mu, -\alpha} c_{\alpha \mu} + {\rm h.c.}\right).
\label{eq:Hamil}
\end{align}
The (pseudo-)spin index $\mu$ and channel index $\alpha$ run over $N$ and $M$ components, respectively.
The $f^2$ state $|\mu \rangle$ forms a channel singlet ($-\alpha$ denotes the counterpart of $\alpha$),
and the $f^1$ state $|\alpha \rangle$ has the energy $E_{\rm ex}$ relative to $|\mu \rangle$.
The Hilbert space of $f$ states is restricted to $|\alpha \rangle$ and $|\mu \rangle$
by using the $X$-operators $X_{\gamma, \gamma'} = |\gamma \rangle \langle \gamma' |$ with $\gamma=\alpha, \mu$, on which $\sum_{\gamma} X_{\gamma, \gamma} = 1$ is imposed.
$c_{\alpha \mu} = N_0^{-1/2} \sum_{\bm{k}} c_{\bm{k} \alpha \mu}$ with $N_0$ being number of sites.
The $M$-channel Coqblin-Schrieffer model is derived from the Hamiltonian~(\ref{eq:Hamil}) as a localized limit $V^2,\ E_{\rm ex} \to \infty$ with $V^2/E_{\rm ex}$ fixed.
%
Exact thermodynamics of the model (\ref{eq:Hamil})\cite{Bolech05} as well as the localized limit\cite{Jerez98} has been derived.

Concerning the dynamical properties, a two-channel case, $N=M=2$, has been clarified
by the numerical renormalization group\cite{Anders05} and by an exact method\cite{Johannesson05}.
General cases have been investigated by
perturbational treatments\cite{Cox-Ruckenstein93, Kroha97, Tsuruta97}.
In this paper, we numerically investigate the dynamical properties of the multichannel Anderson model.
To this end, we develop an algorithm based on the recently developed continuous-time quantum Monte Carlo (CT-QMC) method\cite{Rubtsov05, Werner06, Otsuki-CTQMC}, which is explained in the next section.
We show numerical results for the impurity Green's function and self-energy in \S3.

\section{CT-QMC for the multichannel Anderson model}

\begin{figure}[b]
	\begin{center}
	\includegraphics[width=6cm]{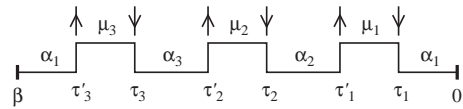}
	\end{center}
	\caption{Diagrammatic representation of a configuration of order $V^6$. 
	The outgoing and incoming allows indicate creation and annihilation of conduction electrons, respectively.}
	\label{fig:mc_config}
\end{figure}

We study the model~(\ref{eq:Hamil}) by the CT-QMC,
which evaluates a perturbation expansion stochastically.
In the present case, we adopt the hybridization expansion\cite{Werner06}.
Since the non-perturbative part is diagonal with respect to $\alpha$ and $\mu$, the efficient algorithm using a `segment' picture is applicable by a slight modification.
Figure~\ref{fig:mc_config} shows a diagram of a configuration of order $V^6$.
Spin states $\mu_i$ and channel states $\alpha_i$ appear alternately, which are hereafter referred to as segment and anti-segment, respectively.
In general, a configuration of order $V^{2k}$ is represented by $q_k \equiv \{ \tau_i, \tau_i', \alpha_i, \mu_i \}$.
The trace over the local states is thus taken into account graphically.
On the other hand, the trace over conduction electrons is evaluated based on Wick's theorem.
A Monte Carlo sampling is performed in the configuration space composed of $k$ and $q_k$.

\begin{figure}[tb]
	\begin{center}
	\includegraphics[width=8cm]{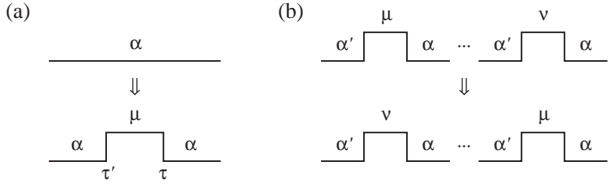}
	\end{center}
	\caption{Update processes: (a) addition of a segment, and (b) exchange of spin indices.}
	\label{fig:update}
\end{figure}

\begin{figure}[tb]
	\begin{center}
	\includegraphics[width=7cm]{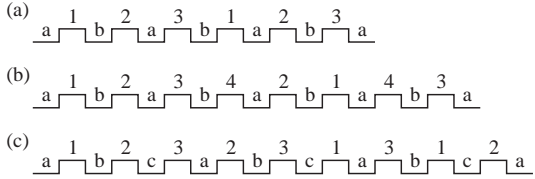}
	\end{center}
	\caption{Examples of diagrams which cannot be reached without the `exchange' process. 
	The spin and channel components are labeled as 1, 2, $\cdots$, and a, b, $\cdots$, respectively.}
	\label{fig:example}
\end{figure}

We perform the following update processes: 
(i) addition/removal of a segment or an anti-segment, and
(ii) exchange of spin or channel indices.
Fig.~\ref{fig:update}(a) shows the addition of a segment.
The index $\mu$ of the segment is randomly chosen, 
and accordingly the update probability differs from that in ref.~\citen{Werner06} by a factor of $N$.
When either $N$ or $M$ is larger than 2, the ergodicity is not satisfied only by process (i).
For example, configurations shown in Fig.~\ref{fig:example} cannot be reached.
This problem can be solved by introducing a process shown in Fig.~\ref{fig:update}(b), which exchanges the spin indices.
We perform a similar update to exchange the channel indices as well.

In the simulation, we observe negative weight configurations for $N=M>2$.
However, since their contribution is less than 10\% in the parameter range shown in this paper, 
the sign problem has little effect on the simulation.

\section{Numerical Results}
In this paper, we restrict ourselves to $N=M$.
We use a rectangular density of states $\rho(\epsilon) = (1/2D) \theta(D-|\epsilon|)$ for conduction electrons with $D=1$.
We fix $NV^2=0.12$ so that the exponent of the Kondo temperature is the same for different $N$.
The width of the localized state $\Delta= \pi V^2 \rho(0)$ is $\Delta \simeq 0.094$ at most (for $N=2$), 
and therefore the effect of finite band width may be neglected.

\subsection{Green's function}
We first show results for the single-particle Green's function $G_f$, 
which is defined in the restricted Hilbert space by
\begin{align}
	G_f({\rm i}\epsilon_n) = -\int_0^{\beta} d\tau
	\langle X_{-\alpha, \mu}(\tau) X_{\mu, -\alpha} \rangle
	{\rm e}^{{\rm i}\epsilon_n \tau},
\end{align}
where $\epsilon_n=(2n+1)\pi T$ is the fermionic Matsubara frequency.
At high frequencies, $G_f$ follows $G_f({\rm i}\epsilon_n) \sim a/{\rm i}\epsilon_n$ with $a<1$, since the Hilbert space is restricted.
The $a$ varies between $1/N$ and $1/M$ depending on $E_{\rm ex}$, and in a special case of $N=M$, $a=1/N$.

\begin{figure}[tb]
	\begin{center}
	\includegraphics[width=8cm]{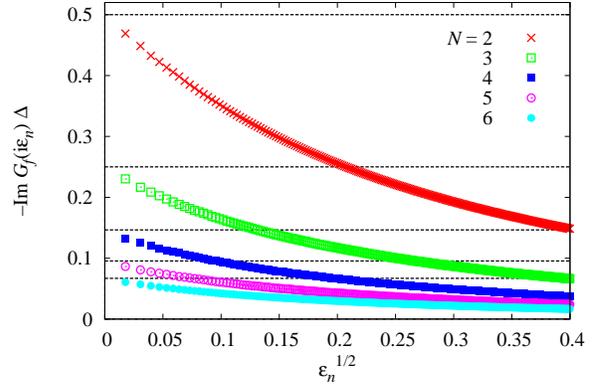}
	\end{center}
	\caption{(Color online) The imaginary part of the Green's function $G_f({\rm i}\epsilon_n)$ for $N=M$, $NV^2 = 0.12$, $E_{\rm ex}=0$ and $T=0.0001$.
	The lines show $\sin^2 ( \pi/2N )$.}
	\label{fig:Gf_omega}
\end{figure}

In Fig.~\ref{fig:Gf_omega}, $-{\rm Im}G_f({\rm i}\epsilon_n) \Delta$ is plotted against $\sqrt{\epsilon_n}$ for $E_{\rm ex}=0$.
For all $N=M$, $G_f$ is expressed as $-{\rm Im}G_f({\rm i}\epsilon_n)=c-b\sqrt{\epsilon_n}$ at low frequencies.
Hence, $G_f(z)$ is non-analytic at $z \to +{\rm i}0$, and the spectrum $-{\rm Im}G_f(\omega+{\rm i}0)$ on real frequencies exhibits a cusp structure expressed by $c-b'|\omega|^{1/2}$,
which has been reported for $N=M=2$\cite{Ludwig-Affleck, Anders05, Johannesson05}.
The value $c$ at $\epsilon_n \to +0$ decreases with increasing $N$.
From Fig.~\ref{fig:Gf_omega} and an analogy with the Friedel sum-rule in the Fermi liquid, we conjecture the following relation:
\begin{align}
	-{\rm Im} G_f(+{\rm i}0) = \frac{1}{\Delta} \sin^2 \left( \frac{\pi}{2N} \right),
\label{eq:Gf-0}
\end{align}
which is indicated in Fig.~\ref{fig:Gf_omega}.
Eq.~(\ref{eq:Gf-0}) includes the result for $N=M=2$\cite{Ludwig-Affleck, Anders05, Johannesson05}, $1/(2\Delta)$, and reduces to the result in the non-crossing approximation\cite{Cox-Ruckenstein93}, $\pi^2/[(N+M)^2 \Delta]$, in the limit $N=M \gg 2$.
For $E_{\rm ex}=0$, the particle-hole symmetry leads to
${\rm Re}G_f({\rm i}\epsilon_n)=0$, 
meaning that the phase shift $\phi$ of conduction electrons at $\omega=0$ is fixed at $\phi=\pi/2$
 irrespective of the value of $N=M$.
Hence, the sine factor in eq.~(\ref{eq:Gf-0}) is not connected with $\phi$ but is due to the imaginary part of the self-energy.

\subsection{Self-Energy}
We discuss the self-energy in the restricted Hilbert space.
The self-energy $\Sigma_f({\rm i}\epsilon_n)$ in the ordinary definition is given by
\begin{align}
	G_f({\rm i}\epsilon_n)
	= \frac{1}{{\rm i}\epsilon_n - \epsilon_f - \Gamma({\rm i}\epsilon_n) - \Sigma_f({\rm i}\epsilon_n)},
\end{align}
where $\Gamma({\rm i}\epsilon_n) = N_0^{-1} \sum_{\bm{k}} V^2 /({\rm i}\epsilon_n - \epsilon_{\bm{k}})$.
In the restricted Hilbert space, $\Sigma_f({\rm i}\epsilon_n)$ diverges according to 
$\Sigma_f({\rm i}\epsilon_n) \sim {\rm i}\epsilon_n (1-1/a)$ at $\epsilon_n \to \infty$, 
since $G_f({\rm i}\epsilon_n) \sim a/{\rm i}\epsilon_n$ with $a<1$.
Although this divergence does not produce any problem with analysis of low-energy properties, it is not convenient in practice.
Thus, we define an alternative self-energy $\tilde{\Sigma}_f$ as follows:
\begin{align}
	G_f({\rm i}\epsilon_n)
	= \frac{a}{{\rm i}\epsilon_n - \tilde{\epsilon}_f - a \Gamma({\rm i}\epsilon_n)
	 - \tilde{\Sigma}_f({\rm i}\epsilon_n)}.
\label{eq:def_self_2}
\end{align}
$\tilde{\Sigma}_f$ is related to the ordinary self-energy $\Sigma_f$ by 
$\tilde{\Sigma}_f = a \Sigma_f + {\rm i}\epsilon_n (1-a) - (\tilde{\epsilon}_f - a\epsilon_f)$, and converges in proportion to $1/{\rm i}\epsilon_n$ at high frequencies.
By using $\tilde{\Sigma}_f$, for example, the renormalization factor $z$ in the Fermi-liquid state is evaluated as
\begin{align}
	z
	&= [ 1- \partial {\rm Im} \Sigma_f({\rm i}\epsilon_n) / \partial \epsilon_n ]_{\epsilon_n \to +0}^{-1} \nonumber \\
	&= a [ 1- \partial {\rm Im} \tilde{\Sigma}_f({\rm i}\epsilon_n) / \partial \epsilon_n ]_{\epsilon_n \to +0}^{-1}
	\equiv a \tilde{z}.
\end{align}
$\tilde{z}$ stands for a quasi-particle weight within the restricted Hilbert space, 
and accordingly $\tilde{\Sigma}_f$ in eq.~(\ref{eq:def_self_2}) may be a reasonable definition.

\begin{figure}[tb]
	\begin{center}
	\includegraphics[width=8cm]{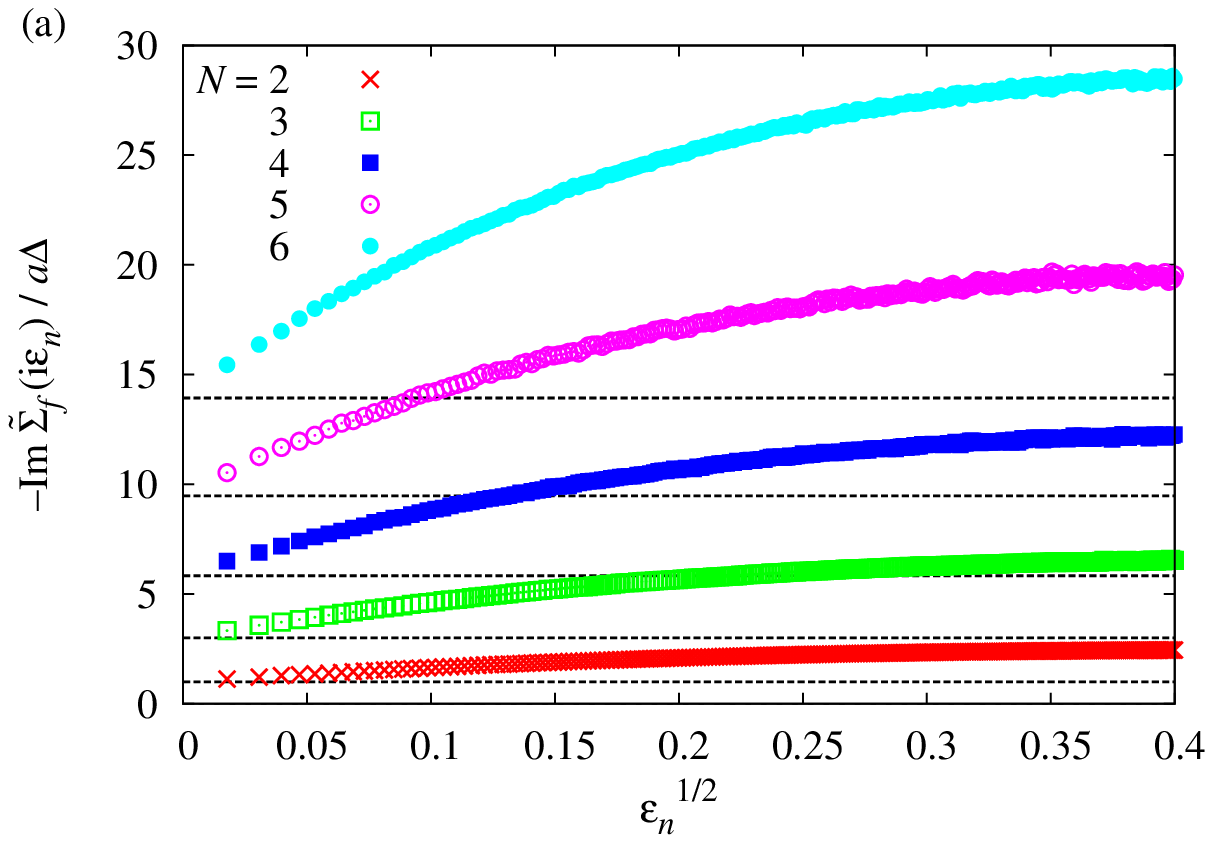}
	\includegraphics[width=8cm]{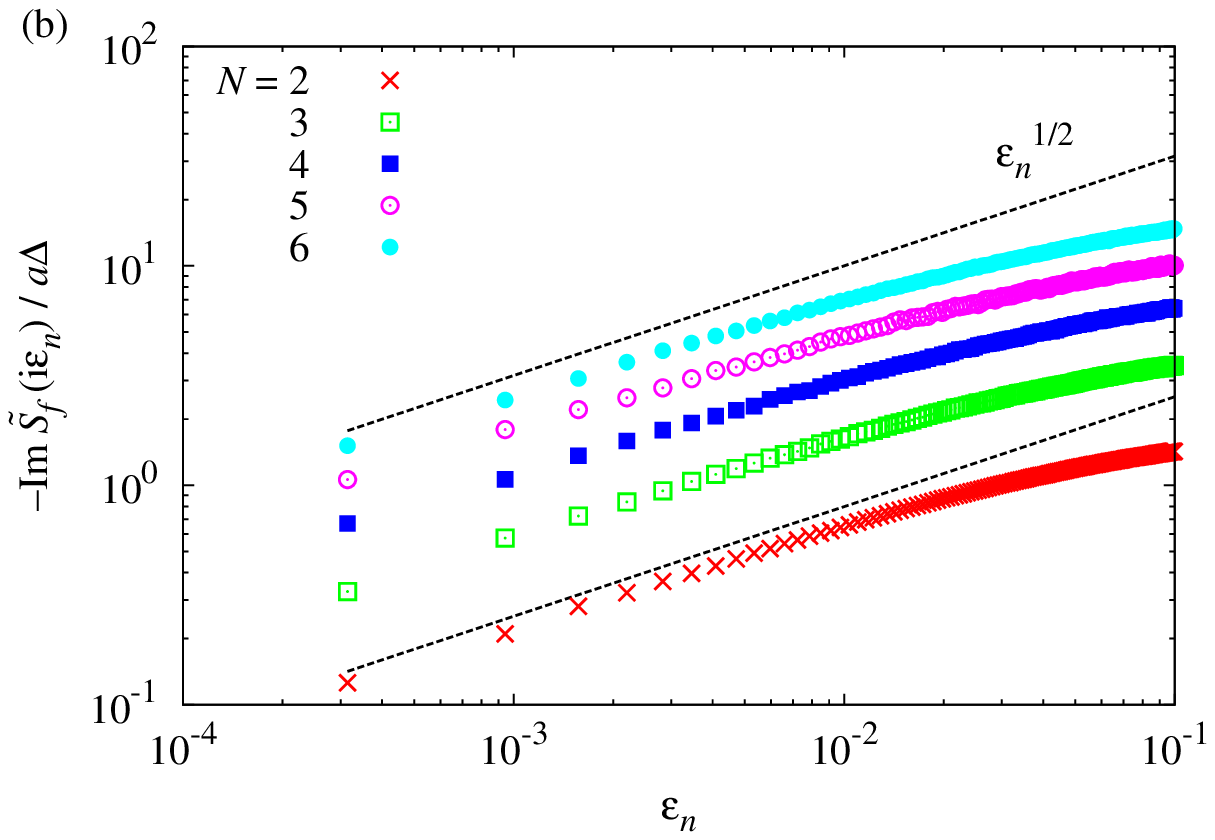}
	\end{center}
	\caption{(Color online) (a) The imaginary part of the self-energy $\tilde{\Sigma}_f({\rm i}\epsilon_n)$ defined by eq.~(\ref{eq:def_self_2}).
	The lines show $\cot^2 ( \pi/2N )$.
	(b) $\tilde{S}_f({\rm i}\epsilon_n) = \tilde{\Sigma}_f({\rm i}\epsilon_n) - \xi a \Gamma({\rm i}\epsilon_n)$ on a log-log scale.
	The parameters are the same as in Fig.~\ref{fig:Gf_omega}.}
	\label{fig:self_omega}
\end{figure}

Figure~\ref{fig:self_omega}(a) shows $-{\rm Im} \tilde{\Sigma}_f({\rm i}\epsilon_n)$ divided by $a\Delta$ as a function of $\sqrt{\epsilon_n}$.
Similarly to $G_f({\rm i}\epsilon_n)$, $\tilde{\Sigma}_f({\rm i}\epsilon_n)$ includes a term proportional to $\sqrt{\epsilon_n}$ in the limit $\epsilon_n \to +0$, 
and converges to a finite value.
To separate the zero-frequency value from $\tilde{\Sigma}_f$, 
we introduce a parameter $\xi$ as follows:
\begin{align}
	\tilde{\Sigma}_f({\rm i}\epsilon_n) &= \xi a \Gamma({\rm i}\epsilon_n) + \tilde{S}_f({\rm i}\epsilon_n).
\label{eq:self-0}
\end{align}
$\xi$ is determined so that 
${\rm Im}\tilde{S}_f(+{\rm i}0) = 0$.
For $N=M$, noting that ${\rm Re}G_f({\rm i}\epsilon_n)=0$, we obtain from eq.~(\ref{eq:Gf-0})
\begin{align}
	\xi = \cot^2 \left( \frac{\pi}{2N} \right).
\label{eq:xi}
\end{align}
In the case of $N=M=2$, 
$\tilde{S}_f({\rm i}\epsilon_n)$ eventually corresponds to the self-energy discussed in refs.~\citen{Anders05} and \citen{Johannesson05}.
Figure~\ref{fig:self_omega}(b) shows $\tilde{S}_f({\rm i}\epsilon_n)$ on a log-log scale.
We can clearly see the power-law behavior
$-{\rm Im}\tilde{S}_f({\rm i}\epsilon_n) \propto |\epsilon_n|^{1/2}$, which means $-{\rm Im}\tilde{S}_f(\omega +{\rm i}0) \propto |\omega|^{1/2}$.

\subsection{Effect of Level Splitting $E_{\rm ex}$: Energy Scale}

So far, we have examined $E_{\rm ex} =0$. We now discuss the effect of $E_{\rm ex}$.
In refs.~\citen{Anders05} and \citen{Johannesson05}, it is reported for $N=M=2$ that ${\rm Im}G_f(+{\rm i}0)$ and ${\rm Im}\tilde{\Sigma}_f(+{\rm i}0)$ do not depend on $E_{\rm ex}$. 
We have confirmed for $N=M \geq 2$ that eqs.~(\ref{eq:Gf-0}) and (\ref{eq:xi}) hold up to $E_{\rm ex}=0.3$ within numerical accuracy.
The finite value of $E_{\rm ex}$ causes an asymmetry of the cusp keeping the value at $\omega=0$: 
$c+b|\omega|^{1/2}$ changes into 
$c+[b_+ \theta(\omega) + b_- \theta(-\omega) ]|\omega|^{1/2}$.

\begin{figure}[t]
	\begin{center}
	\includegraphics[width=7.5cm]{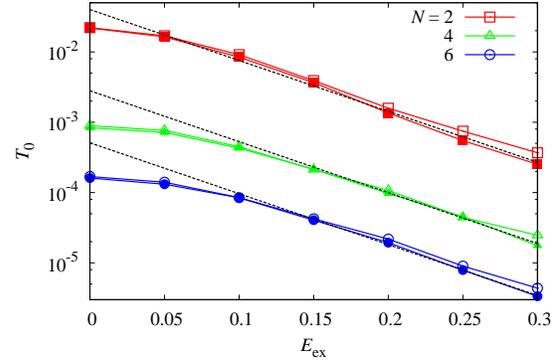}
	\end{center}
	\caption{(Color online) A characteristic energy scale $T_0$ defined in eq.~(\ref{eq:energy_scale}) as a function of $E_{\rm ex}$. $T_0$ is evaluated from $\tilde{S}_f({\rm i}\epsilon_0)$ at $T=0.0005$ (open symbols) and $T=0.00025$ (closed symbols).
	The lines show $\exp[-E_{\rm ex}/N V^2 \rho(0)]$.}
	\label{fig:energy_scale}
\end{figure}

As $E_{\rm ex}$ increases, the energy scale becomes smaller.
We define a characteristic energy scale $T_0$ in terms of $\tilde{S}_f$ by
\begin{align}
	-{\rm Im} \tilde{S}_f({\rm i}\epsilon_n) / a\Delta 
	\sim ( \epsilon_n / T_0 )^{1/2},
\label{eq:energy_scale}
\end{align}
in the limit $\epsilon_n \to 0$.
Because $T_0$ may be defined with an arbitrary factor,
we shall discuss only its exponent.
In Fig.~\ref{fig:energy_scale}, we show $T_0$ as a function of $E_{\rm ex}$.
$T_0$ follows $T_0 \propto T_{\rm K} \propto \exp(-1/g)$ with $g=N V^2 \rho(0)/E_{\rm ex}$ for $E_{\rm ex} \gtrsim 0.15$, namely $g \lesssim 0.4$.
We conclude that the exponent of the energy scale of the non-Fermi liquid self-energy agrees with the Kondo temperature $T_{\rm K}$ in the corresponding single-channel model.

\section{Summary}
We have presented the impurity Green's function $G_f({\rm i}\epsilon_n)$ and the self-energy $\tilde{\Sigma}_f({\rm i}\epsilon_n)$ in the non-Fermi liquid state using the CT-QMC extended to the multichannel Anderson model.
For $N=M$, $G_f$ and $\tilde{\Sigma}_f$ are non-analytic at $\omega=0$ as $|\omega|^{1/2}$.
The zero-frequency spectrum ${\rm Im}G_f(+{\rm i}0)$ does not depend on the excitation energy $E_{\rm ex}$, and 
correspondingly ${\rm Im}\tilde{\Sigma}_f(+{\rm i}0)$ has a finite value.
These values depend only on $N$ ($=M$), and seem to be expressed as eqs.~(\ref{eq:Gf-0}), (\ref{eq:self-0}) and (\ref{eq:xi}).
An analysis of general $N$, $M$ is left for future work.

We acknowledge Prof. Y. Kuramoto for comments on the manuscript.
This work was supported by a Grant-in-Aid for Scientific Research on Innovative Areas ``Heavy Electrons" (No. 20102008) of the Ministry of Education, Culture, Sports, Science, and Technology, Japan.

\end{document}